\documentclass[a4paper,11pt]{article}
\usepackage{graphicx}                
\usepackage{amssymb}
\usepackage{amsmath}
 
\newcommand{\be}{\begin{equation}} \newcommand{\ee}{\end{equation}}
\newcommand{\ba}{\begin{eqnarray}} \newcommand{\ea}{\end{eqnarray}}
\newcommand{\ban}{\begin{eqnarray*}}
\newcommand{\ean}{\end{eqnarray*}}
\newcommand{\barr}{\be\left\{\begin{array}}
\newcommand{\earr}{\end{array}\right.\ee}

\title{Causal Relativistic Fluid Dynamics} 

\author{X. Chen,\\
E. A Spiegel$^a$\\\\
$^a$ Dept. of Astronomy, Columbia University, New York, NY, USA
}

\begin{document}
\maketitle




\begin{abstract}
\noindent
We derive causal relativistic fluid dynamical equations from the relaxation model of kinetic theory as in a procedure previously 
applied in the case of non-relativistic rarefied gases in Ref.~\cite{bgk1}.  By treating space and time on an equal footing and avoiding the
iterative steps of the conventional Chapman-Enskog --- CE---method, we are able to derive causal equations
in the first order of the expansion in terms of the mean flight time of the particles.  This is in contrast to what is found using the CE approach.  We illustrate the general results with the example of
a gas of identical ultrarelativistic particles such as photons
under the assumptions of homogeneity and isotropy.  When we couple the fluid dynamical equations to Einstein's equation we find, in addition to the geometry-driven expanding solution of the FRW model, a second, matter-driven nonequilibrium solution to the equations.  In only the second solution, entropy is produced at a significant rate. 
\end{abstract}

\section{Introduction}
 
Recent developments in the study of very hot objects, especially of the early universe, have underscored the need for  an adequate description of relativistic fluid dynamics.    Yet  there are 
questions about the derivation of the basic equations from kinetic theory that are not completely
settled.   It may seem surprising that any doubt should arise in 
regard to fluid dynamical equations whose provenance stretches  back into the nineteenth century, but the application of fluid dynamics under extreme conditions may well require a reexamination of the foundations.     And the answers to the questions raised may have interesting implications especially in
the relativistic case.  An example that has led to some discussion is the suggestion that bulk viscosity (the dissipative resistance to changes in volume) may give modify the expansion rate in a 
self-gravitating gas \cite{zim96,maa95,mak98,dur08}.    We shall not delve into that controversial issue, but we will use the relativistic expansion of a self-gravitating gas to illustrate the content of the equations derived here.  However, our main aim in this work is the derivation of causal equations for relativistic fluid dynamics.

The failure of causality that was encountered in early derivations of the relativistic fluid equations from kinetic theory was foreshadowed in the nonrelativistic version of those equations \cite{mul67,isr76}.  Further indications that all was not well were
revealed when measurements on the propagation of sound in rarefied gases failed to confirm the theoretical predictions based on the Navier-Stokes equations \cite{uhl63}.   Those equations are commonly derived by the Chapman-Enskog procedure 
in the classical case \cite{cha60}.   The experiments helped to identify a weakness of the standard derivations of the fluid equations \cite{bgk1} that make use of an iterative procedure
whose ill effects are most pronounced when the mean flight
times of particles are not sufficiently short.  The approximation 
that we advocate here allows for the possibility of long mean
flight times and the consequent deviations from equilibrium.
  
Fluid dynamical equations may be derived as the the three lowest moments over momentum space of the equations of kinetic theory,  such as Boltzmann's equation.   However, 
the stress tensor, which appears in those moment equations, is 
not prescribed and an additional equation for it is needed to complete the fluid description.   A similar limitation occurs for the third moment of the kinetic equation --- the heat equation ---  which contains no specification of the heat flux.  An equation for
that moment likewise must be prescribed before the system of continuum equations is complete.  

A standard way to prescribe the needed additional equations is the Chapman-Enskog approach in which an approximate solution to the kinetic equation is found by expanding the one-particle distribution function in the mean free path or mean flight time of the constituent particles.    This approximation is then used to obtain expressions for the pressure tensor and heat flux that may close the system of continuum equations.

It has been the practice, following Chapman and Enskog,  to simplify the derivations of expressions for the stress tensor at any order by introducing results found in previous orders.   In the nonrelativistic  case this is done in such a way that, as Uhlenbeck \cite{uhl63} has described it, ``the successive hydrodynamical equations are {\it always} of the {\it first} order in time derivatives  of the macroscopic variables, but are successively of {\it higher order} in the space derivatives" (italics his).    This leads to parabolic
equations and that is at the origin of the failure of causality in both the classical and the relativistic cases \cite{isr63} since space and time are not kept on the same footing in this approach.

Grad \cite{gra63} has observed that  ``the Chapman-Enskog expansion is asymptotic rather than convergent.''    To us, Grad's
suggestion implies that the C-E iteration is not appropriate when the expansion parameter is not truly infinitesimal.   We have found support for this suggestion in the non-relativistic case where fluid equations derived from kinetic theory by omitting the iterative step of Chapman and Enskog give better agreement with experiment for the phase speed of ultrasound and of the structures of shocks than do the Navier-Stokes equations \cite{bgk1,thi03,dell07}, when the
mean free paths are long compared to macroscopic scales.

Though not all difficulties have been resolved as yet \cite{thi03}, it appears worth treating the relativistic case without introducing the iterative approximation of Chapman and Enskog as we have done already for the case of radiative radiative fluid dynamics 
\cite{cherad, chebulk}.   Here, we emphasize the formal structure
of the derivations that lead to causal, covariant equations of fluid dynamics in the first order in the expansion in terms of the mean flight time of particles (to be denoted herein as $\epsilon$). 
In the relativistic case, many (if not most) derivations of the fluid dynamics equations from kinetic theory follow the procedure of Chapman and Enskog \cite{uhl63}.   

One result of applying the C-E procedure is that it leads in the first order of the expansions to non-causal equations that indicate (unphysical) instability for the equilibrium state variables \cite{mul67,isr76,isr79a,pav82,his83, maa95}.  In previous work the difficulty has been circumvented by going to a second-order theory phenomenologically by M\"uller \cite{mul67} in the classical case and later extended to the relativistic case both phenomenologically
and in kinetic theory by Israel and others in an ever developing literature beginning, for example, in \cite{isr76, isr79a, pav82}.     Also noteworthy is the version of Chapman-Enskoggery formulated by Van Kampen \cite{kam87} who saw the development in terms of a singular perturbation theory based on the two-time approach.  The meanings of fast and slow times that the scheme involves is not invariant under change of frame and the fluid equations found by Van Kampen are consequently not Lorentz covariant. 
  
In the derivation given below, we avoid the C-E iteration and so obtain relativistic fluid equations that are both covariant and causal
in the first order of the development in mean flight time (or mean
free path).  For the purpose, we use the relativistic form of the relaxation model of kinetic theory, which is more general, if less explicit, than the relativistic form of Boltzmann's kinetic equation.  Moreover, the derivations and their meaning are most easily
appreciated in the simpler case of the relaxation model of kinetic theory.  
\section{The Kinetic Model}
As our aim is to exhibit a procedure for deriving causal fluid equations, we shall consider only the relatively simple example of 
a gas of particles of a single kind having no internal degrees of freedom.  Although the typical examples in which relativistic fluids
are studied are astrophysical plasmas that are rich soups of 
particle species, we make this simplification in order to be able 
to carry out our procedure in a reasonably transparent manner.   This approach may be appropriate when there is a dominant constituent in the mix of particles, and that is the view we adopt in the example discussed in the latter portion of this paper.  

We shall work (at first only) in Minkowski spacetime with coordinates $x^\mu$ and adopt units in which the speed of light is unity.  Each particle in our gas of identical particles has rest mass $m$ (which may be zero).  We describe the state of the gas with a scalar one-particle distribution function, $f(x^\mu, p^\nu)$, where $p^\nu$ is the four momentum of a particle.    From the equation for $f$ we wish to derive equations for the macroscopic fluid quantities including the local macroscopic fluid velocity, 
$u^\mu$.   Choosing a $u^\mu$ is tantamount to specifying (what astronomers call) the local standard of rest, that is, a local rest frame.   For ease of presentation, we postpone the prescription of $u^\mu$ until we have developed the subject further.

We describe the evolution of $f$ by an equation of the form \be
 p^\mu f_{,\mu}=\alpha - \varkappa f \ .
\label{1} \ee 
The quantity $\alpha$
is the rate at which particles are scattered into states with 
momentum $p^\mu$ from some other momenta; in that sense, 
it is like a transition probability.   The second term on the right of 
(\ref{1}) is the rate at which particles are being scattered out of the momentum state $p^\nu$ with $\varkappa$ being
proportional to the inverse of the mean flight time of particles.
 The same equation has been used to study radiative transfer and, in the relativistic case, was discussed by L.H. Thomas \cite{tho30, sim63, and72} who gave the transformation rules 
under frame changes for $f$, $\alpha$ and 
$\varkappa$.  Boltzmann's equation, which is based on two-particle interactions, can be written in this form with 
$\alpha$ as a quadratic functional of $f$ and $\varkappa$ a linear functional of $f$.
  
We presume the existence of a local equilibrium, $f_{(0)}$,
for which the right hand side of (\ref{1})) vanishes.  This property
leads to the relation \be
f_{(0)} = \frac{\alpha}{\varkappa} \ , \label{KP} \ee
which is analogous to the Kirchhoff-Planck law of radiation theory. 
We adopt this expression for $f_{(0)}$ with the understanding that it represents only a {\em local} equilibrium in which the macroscopic quantities it depends upon 
(pressure, temperature and so on) may themselves vary with position and time.   Then (\ref{1}) becomes the relaxation model \be
 p^\mu f_{,\mu}=\varkappa (f_{(0)}-f) \ .
\label{relax} \ee 
 
The nonrelativistic version of (\ref{relax}) was adopted for the study of material particles by Bhatnagar,
Gross and Krook \cite{bgk54} and by Welander \cite{wel54}.  It is also known as the BGK
model and, more recently, as the WKBG model.   Krook had been a student of Eddington and
he, at least, may well have had the radiative analogue in mind in 
thinking about this problem.  In our work, we have benefitted from the formal equivalence of the various versions
of the theory.  In particular, we learn from Thomas' discussion of the relativistic transfer equation that
$\varkappa$ may be expressed in terms of its value in the local rest frame by the transformation rule
\be 
\varkappa=u^\mu p_\mu \hat\varkappa  \label{WBGK} \ee
where $\hat\varkappa$ is the inverse of the mean flight time
evaluated in the local rest frame of the fluid.    A hat placed on a symbol
will denote evaluation of a quantity in the local
rest frame throughout.

The equation governing the relativistic relaxation model is then \be
\epsilon p^\mu f_{,\mu}=u_\nu\,p^\nu(f_{(0)} -  f)\; \label{bgk1} \ee
where $\epsilon = 1/\hat\varkappa_0\; .$  In our formulation of the
equation, $f$ is a scalar so that (\ref{bgk1}) is manifestly
covariant in keeping  with the practice of much of modern relativistic radiative transfer theory \cite{haz59}.  In the present study, based 
on the relaxation model, we shall not allow for an explicit dependence of $\epsilon$ on momentum and will take it to be a function of only the local macroscopic properties of the gas.  

In previous work with the relativistic relaxation model, Marle 
\cite{mar69} wrote the relativistic relaxation model as in (\ref{relax}) as if $\varkappa$ were an invariant; he did not discuss the issue of its  transformation properties.   Though Anderson and Witting 
\cite{and74a,and74b} also eluded the issue of transformation properties, they did introduce the factor $u_\nu\,p^\nu$.  Hence, if the $1/\tau$ in their treatment that apparently corresponds to that of Marle (and to our $\varkappa$) were replaced by the equivalent of 
$\hat\varkappa$, their formulation would be covariant.  
\section{Macroscopic Description}
In the Boltzmann version of the kinetic equation (with no quantum
mechanics),
the right hand side describes the results of elastic, two-particle collisions.  These
conserve five quantities written in the relativistic case as
$\psi^A$, $A=(0,1,...,4)=(\mu,4)$, where $\psi^4=1$ and $\psi^\mu=p^\mu$.  That is, multiplication of (\ref{bgk1}) by $\psi^A$ followed by integration over momentum space gives zero on the right side
and the resulting equations conserve the number, momenta and energies of the colliding particles.
In the case of the relaxation model, we {\it impose} the same 
conservation conditions, by requiring that multiplication by
$\psi^A$ followed by integration over momentum space gives zero on the right hand side.   Since we have assumed that $\epsilon$ is independent of $p^\mu$, this condition, known as the matching condition \cite{ste72}, may be written as \be
\int \psi^A f dP = \int \psi^A f_{(0)} dP\, . \label{match} \ee
Here $dP=d^3p/e$ is the invariant volume element in momentum space                                                                                                                  where $d^3p$ is the three-dimensional volume element in momentum space \cite{ste72,lan87} and $e$ is the
particle energy.  (This result is discussed in terms of the geometrical basis of transport theory in \cite{lind66}.)

The matching condition (\ref{match})
leads to an osculating equilibrium distribution function for which the
fundamental invariants have the same values as in the actual state of the gas.   We may then obtain the macroscopic equations, by multiplying equation (\ref{1}) by $\psi^A$ and integrating over momentum space.   
Before proceeding, we should recall that we are working just now in flat Minkowski spacetime.  Hence we do not yet need to distinguish between partial and covariant derivatives.   To go to curved space or even  to introduce curvilinear coordinates, we replace the partial derivatives by covariant derivatives as appropriate and signify this by replacement of commas in the subscripts by semicolons.  This replacement may be justified formally by including a term  
$\dot p^\mu \, \partial f/\partial p^\mu$
on the left side of  the kinetic equation and complementing it with the geodesic equation 
for $\dot p^\mu$ where the overdot denotes differentiation with respect to proper time (or to an appropriate path parameter).  The commas and semicolons then take care of themselves.

When we multiply (\ref{1}) by $\psi^4=1$, we find that \be
\int p^\mu f_{,\mu}dP=0. \ee
Since $p^\mu$ is independent of $x^\mu$, this leads to \be 
N^\mu_{\ ,\mu}=0 \label{7} \ee 
where   
\be N^\mu=\int p^\mu fdP\; \ .
\label{8} \ee 
Then, when we multiply by $\psi^\mu=p^\mu$ and integrate, we find \be
T^{\mu\nu}_{\ \ \, , \mu}=0 \label{11} \ee 
where \be T^{\mu\nu}=\int p^\mu p^\nu fdP \ .
\label{12} \ee 

It can be helpful to decompose these
basic quantities.  For that purpose,  we employ the projection operator \be
h^{\mu \nu} = g^{\mu \nu} - u^\mu u^\nu \label{h} \ee
where $ g^{\mu \nu}$ is the metric tensor (for now, the Minkowski metric).
The component of $N^\mu$ along $u^\mu$ may be identified as the number density of particles
in the rest frame, \be
\hat N = u_\mu N^\mu \ ,  \ee  
and its spacelike component (orthogonal to $u^\mu$) is \be 
J^\mu = h^{\mu \nu} N_\nu \ . \ee

To separate $T^{\mu\nu}$ into its  constituent parts we use the (Eckart \cite{eck40}) decomposition of $p^\mu = (e, {\bf p})$
where  ${\bf p}$ is the three-momentum and we write $p=|{\bf p}|$.  
We let $\hat e = u_\mu p^\mu$
be the particle energy in the local rest frame of the fluid with a corresponding meaning for $\hat p$.  We further introduce the 
spacelike vector, $l^\mu$, with the properties \be
l^\mu l_\mu = -1 \qquad \qquad{\rm and}\qquad \qquad l^\mu u_\mu = 0.\ee
Then we may write \be
p^\mu = \hat e u^\mu + \hat p\, l^\mu  \label{split} \ee
and note that \be
l^\mu l_\mu = \frac{h^\rho_{\ \sigma}\,  p^\sigma p_\rho}{{\hat p^2}} = \frac{m^2-\hat e^2}{\hat p^2} = -1.\ee
These considerations apply also for a gas of photons, in which case, $p=e$ and $m=0$.

The stress tensor may then be written in component form as  \be
T^{\mu \nu} = E u^\mu u^\nu +  F^\mu u^\nu + F^\nu u^\mu + 
P^{\mu \nu}   \  ,  \label{prtens} \ee
where  \be
E = \int \hat e^2fdP \qquad
F^\mu =  \int \hat e \hat p l^\mu f dP   \qquad P^{\mu \nu}= \int \hat p^2 l^\mu l^\nu fdP
 \ . \label{comp} \end{equation}
And, of course,   \be
E= u^\mu u^\nu T_{\mu \nu} \qquad
F^\mu =  h^{\mu \nu} u^\rho T_{\nu \rho} \qquad P^{\mu \nu}= h^{\mu \rho} h^{\nu \sigma} T_{\rho \sigma}
\ . \label{decomps}
\end{equation}
(Hats are omitted for cosmetic reasons.) 
Then (\ref{7}) and (\ref{11}) are written out as
\ba & & u^\mu N_{,\mu}+N\vartheta+J^\mu_{\ ,\mu} = 0 \label{cont} \\ & &
u^\mu(Eu^\nu)_{,\mu} + E u^\nu\vartheta +(F^\mu u^\nu+F^\nu
u^\mu)_{,\mu}+P^{\mu\nu}_{\ \ ,\mu}=0 \label{mo} \ea 
where $\vartheta=u^\mu_{\ ,\mu}$.
A few more elementary steps are needed to complete this formulation of the basic equations, beginning
with the specification of the macroscopic velocity.
\section{Frame Choices}
Equations (\ref{cont}) and (\ref{mo}) provide relations among the macroscopic quantities
$N$, $E$, $u^\mu$, $F^\mu$, $J^\mu$ and $P^{\mu\nu}$.  
We have yet to choose the frame in which to express these relations.  That is, we need to specify 
$u^\mu$, which van Kampen \cite{kam68} regards as a thermodynamic variable in its own right.  And
we would do well to choose a frame that simplifies the equations.  
Indeed, any five relations among the macroscopic variables specifies a frame, so many choices are possible, the two most widely considered being those proposed by Landau and Lifshitz 
\cite{lan87} and by Eckart \cite{eck40}.

\begin{enumerate}

\item {\sl The Landau-Lifshitz Conditions} \\ 
Landau and Lifshitz specify a preferred frame by imposing these five relations among the macroscopic quantities:
\be u_\mu N^\mu=u_\mu N_{(0)}^\mu
\label{19}
\ee and \be u_\mu T^{\mu\nu}=u_\mu T_{(0)}^{\mu\nu}
\label{20}
\ee where the subscript ${(0)}$ implies evaluation in local equilibrium; $N_{(0)}^\mu$ and $T_{(0)}^{\mu\nu}$ are the appropriate moments of $f_{(0)}$. 
 
In local equilibrium, to good approximation, the number 
current and the energy flux each vanish.  Equations (\ref{19}) and (\ref{20}) then lead to 
\ba N & \equiv & N_{(0)} \nonumber \\ E & \equiv & E_{(0)} \label{21} \\
F^\mu & \equiv & 0 \nonumber \ea 
Thus, in going to the Landau-Lifshitz frame we transform away the energy flux and reduce (\ref{mo}) to 
\ba & & u^\mu N_{,\mu}+N\vartheta+J^\mu_{\ ,\mu} = 0 \label{LL1} \\ & &
u^\mu(Eu^\nu)_{,\mu} + E u^\nu\vartheta +P^{\mu\nu}_{\ \ ,\mu}=0. \label{LL2} \ea 

With $(\ )^\cdot = u^\mu(\ )_{, \mu}$
the second of these equations may be written as \be 
Eu^\mu \vartheta+(Eu^\mu)^\cdot+P^{\mu\nu}_{\ \ ,\nu} =0 .
\label{27}\ee  
It can then be broken down into components.

On projecting (\ref{27}) in the direction of $u_\mu$, we get the
energy conservation equation \be 
\dot E + E\vartheta +  u_\mu P^{\mu\nu}_{\ \ ,\nu}=0\; .
\label{28}
\ee 
When we project (\ref{27}) with the operator 
$h_{\rho\mu}$ of (\ref{h}), 
we obtain
\be E\dot u^\mu+P^{\mu\nu}_{\ \  ,\nu}=0\; ,
\label{29} \ee 
which is the equation of motion.    

We shall use the Landau-Lifshitz frame choice here but it is also worth recording another much-favored choice, that of Eckart. 

\item {\sl The Eckart Matching Conditions} \\ 
In the frame chosen by Eckart, the matching
conditions are given by $N^\mu=N_{(0)}^\mu$ and $u_\mu u_\nu
T^{\mu\nu}=u_\mu u_\nu T_{(0)}^{\mu\nu}$. These conditions lead to 
\ba N & \equiv & N_{(0)} \nonumber \\ E
& \equiv & E_{(0)} \label{24} \\ J^\mu & \equiv & 0 \; .\nonumber \ea  Hence, in the Eckart frame, the particle current is entirely convective.

In this frame, the fluid dynamical equations are 
\ba & & u^\mu N_{,\mu}+N\vartheta = 0 \label{E1} \\ & &
u^\mu(Eu^\nu)_{,\mu} + E u^\nu\vartheta + (F^\mu u^\nu +F^\nu
u^\mu)_{,\mu} + P^{\mu\nu}_{\ \ ,\mu}=0\; . \label{E2} \ea 
\end{enumerate}
For either of these two choices of frame, as for any other one that is consistent with the physics,
we thus have, to first order accuracy in $\epsilon$, causal, covariant fluid equations.
However, these descriptions are incomplete until a means of computing 
$P^{\mu \nu}$ has been provided.   A standard procedure for this is to try to relate $P^{\mu \nu}$ to the lower moments of $f$.   
Among the ways for finding that relation, the most direct is perhaps to derive an 
approximate solution to the kinetic equation for $f$ and, from that, to compute the moments 
of $f$ up to $P^{\mu \nu}$ so as to close the system.
In our procedure, we enlarge the usual notion of closure relations to allow a dependence on the derivatives of the lower moments as well.  The appearance of derivatives in our first-order expression for  the pressure tensor may be thought of as a process-dependent feature of our approximation.  Our aim here is to illustrate this point and, with that intention, we next derive an expression for $P^{\mu\nu}$ in the case of a simple situation of astrophysical interest. 
\section{Approximating $f$}
\subsection{An Illustrative Example}
To illustrate our approach to the derivation of a closure relation without invoking the C-E iteration, we continue to consider the 
case of a gas consisting of one type of particle with no
internal degrees of freedom.    To simplify the presentation even
further, we presume that the particles are ultrarelativistic so
that their masses may be left out of account in a description that
would be suitable for a  photon-dominated medium.   This approach leads to a closure approximation that is representative of more general situations without the necessity of performing the more arduous calculations that situations with complicated particle mixes would call for.  And even in the case of the photon gas, interactions are possible without the intervention of real massive particles, since
very energetic photons may scatter off one another
by creating virtual $e^+\negthinspace\negthinspace-\negthinspace e^-$ pairs.  
 
The commonality of this example with studies of radiative transfer
in material media is also helpful in other ways,  as we have already mentioned \cite{tho30}.  Just as for photons, we will not assume that 
the particle number is conserved (though we are not 
explicitly including any quantum mechanics).  Hence 
we need consider simply 
\be T^{\mu\nu}_{\ \ ,\mu}=0\, , \ee
to which our fluid dynamical equations boil down.   
Then, to complete the formulation, we complement equations (\ref{28}) and (\ref{29}) with an expression for $P^{\mu\nu}$
in terms of $E$ and $\dot E$.   

To find a closure relation for $P^{\mu\nu}$ we
introduce a series expansion of $f$ in terms of $\epsilon$ into (\ref{bgk1}) to find an approximate solution.  Then we use that to evaluate and relate the relevant moments of $f$.   
For conventional radiative fluid dynamics, we have carried this 
procedure out in~\cite{cherad}.   There we took $u^\mu$ as
the velocity of the ambient medium and treated it as known.
In the present example, there is no background medium and we 
are working with a gas of ultrarelativistic particles whose masses 
we neglect as for a photon 
gas.  We now have in mind a $u^\mu$ that is the appropriate velocity field of the fluid itself even in the case of a photon fluid.    Although we then obtain a pressure tensor of the same form as in \cite{cherad}, the change in viewpoint makes it desirable to provide a sketch of the derivation since the closure relation obtained is central to the present work\footnote{Treating the photon gas as a fluid in its own right by working in the radiation's inherent reference frame is also a useful way to obtain approximate solutions in conventional radiative transfer theory.  The relationship of that approach to the more familiar Eddington factor method will be described elsewhere.}.
   
A nice simplification of the ultrarelativistic problem is that a
particle's energy is equal to the magnitude
of its three-momentum.  That is, $e = p := |{\bf p}|$ (with $c=1$)
where $p^\mu =  (e, {\bf p})$.     Then, in thinking about the macroscopic aspects, it is useful to
introduce the null vector $n^\mu$ such that $n^\mu \hat e = p^\mu$ and, as in (\ref{split}), 
to decompose it into \be
n^\mu = u^\mu + l^\mu .\ee
As before, \be
l^\mu l\mu = -1 \qquad \qquad {\rm and} \qquad \qquad l^\mu u_\mu = 0 \; . \ee
Thus,  as in (\ref{prtens}), the components of the stress tensor may be expressed in terms of the quantities defined in (\ref{comp}). 
\subsection{The expansion of $f$}
To develop an expression for $P^{\mu \nu}$ by seeking an approximation for $f$, we return to the relaxation model 
(\ref{bgk1}).  Since $ p^\mu=n^\mu \hat e $ and 
$u^\mu p_\mu = \hat e$, it may be written as \be 
\epsilon n^\mu f_{, \mu} = f_{(0)} - f \ ,\label{min} \ee
where, in the present instance, $p_\mu p^\mu = 0$.
We then seek approximate solutions to this equation in the form \be
f = f_{(0)} + \epsilon f_1 + {\cal O}(\epsilon^2) \;  . \label{exp} \ee
Though we shall not always indicate the ${\cal O}(\epsilon^2)$ error terms, they are important and should be kept in mind.   Then, 
from (\ref{min}), we see that \be
f_1 = - n^\mu f_{(0),\mu} \ . \ee
With this approximation, we may evaluate the stress tensor but first we must specify $f_{(0)}$.

For the photon gas of the present example, the equilibrium distribution \be
f_{(0)}=[e^{\beta u^\mu p_\mu}-1]^{-1}
\label{30}
\ee
is appropriate.  This is the equilibrium solution for a gas of bosons with zero  chemical potential.
As usual, we use this expression in a local sense and allow that 
both $\beta$ ($=1/T$ in suitable units) 
and $u^\mu$ may depend weakly (to use this dangerous term) on $x^\mu$.   

The relaxation equation is always dragging $f$ toward the evolving equilibrium whose changes are connected to
$f$ itself through the matching condition, (\ref{match}).  The
variation of $f_{(0)}$ with $x^\mu$ is must be found and are not
specified {\it a priori}; the changes take place through the dependence of $f_{(0)}$ on $\beta$ and $u^\mu$ each of which may itself depend upon $x^\mu$.  We take advantage of  this structure of $f_{(0)}$ by using the chain rule to write \be 
\partial_\mu f_{(0)} = \left(T_{,\mu} \partial_T + \hat e_{,\mu} \partial_{\hat e}\right) f_{(0)} \label{chain} \ee
where $\partial_z$ means a partial derivative with respect to $z$.   In higher orders, we would need
to include derivatives with respect to $n^\mu$ and any nonconstancy of $\epsilon$ would also make
itself felt (just as in second order in the nonrelativistic case 
\cite{thi03}).   For particles with nonzero rest mass, we might need to include a chemical potential as well.  

A further  simplification is that, as for the photon gas, $f_{(0)}$ depends on $\hat e$ and $T$ only through the ratio
$\hat e/T$  (Stefan's law) so that we may write \be
\partial_\mu f_{(0)}= \left(T_{,\mu} - T l^\rho u_{\rho,\mu}\right) \partial_T f_{(0)} \ . \ee
And so we conclude that, to first order in $\epsilon$, \be
f = f_{(0)} - \epsilon n^\mu  \left(T_{,\mu} - T l^\rho u_{\rho,\mu}\right) \partial_T f_{(0)} + {\cal O}(\epsilon^2) \; .
\label{1storder} \ee
\section{The Closure Approximation}
\subsection{Zeroth order}
In leading order, with $f=f_{(0)}$, we may use the last of  
(\ref{comp}) to evaluate the pressure tensor, which becomes  \be
P^{\mu \nu}  = \int \hat p^2l^\mu l^\nu f_{(0)} dP .
\ee
Because the medium is locally isotropic in zeroth order, the angular integral in momentum space involves only products of the $l^\mu$.  Such integrals are 
evaluated by expressing the results in terms of $u^\mu$ and
$g^{\mu\nu}$ and making suitable choices for the coefficients.  We find that \be
\int l^\mu l^\nu d\Omega = -  \frac {4\pi}{3} h^{\mu \nu} \ .\ee
We then obtain (in zeroth order) the closure relation \be
P^{\mu \nu}  =   - Ph^{\mu\nu} \ee  
where  \be
P =  \frac{4\pi}{3} \int \frac{\hat p^4}{\hat e} f_{(0)} d\hat p \ee 
is identified as the pressure.
On using the first of  (\ref{comp}), we see that \be
P = \frac{1}{3}E\ , \label{state} \ee
which is the usual equation of state for equilibrium radiation,
where $E$ is the energy density.  Moreover, for the equilibrium 
(\ref{30}) we know
that $E=aT^4$, where $a$ is the radiation constant, so that serves to tie in the temperature. 
We may also obtain the equations of motion for a perfect fluid 
at this order but we have no need of them
and do not exhibit them here. 
\subsubsection{First order}
Next we introduce (\ref{1storder}) into (\ref{comp}).  Performing the angle integral is again straightforward and we find that \be
\int l^\mu l^\nu l^\rho l^\sigma d\Omega = \frac{4\pi}{15}\left( h^{\mu \nu} h^{\rho \sigma} 
+ h^{\mu \rho} h^{\nu \sigma} + h^{\mu \sigma} h^{\nu \rho}\right) \ . \ee
So we obtain on using the various definitions that \be
P^{\mu\nu}=- P h^{\mu\nu} + 
h^{\mu\nu}\mu \left(\frac{\dot E}{E}+
\frac{4}{3}\vartheta\right)+\Xi^{\mu\nu}   \label{30.5}  \ee 
in which \be
\Xi^{\mu\nu}=\frac{4\mu}{5}\tau^{\mu\nu\rho\sigma}u_{\rho,\sigma}
\label{31}
\ee 
is the viscous shear stress tensor where  \begin{equation}
\tau^{\mu\nu\rho\sigma}=h^{\mu\rho}h^{\nu\sigma}+
h^{\mu\sigma}h^{\nu\rho}-\frac{2}{3}h^{\mu\nu}h^{\rho\sigma} 
\label{tau} \end{equation}
and $\mu = P\epsilon$ is  the viscosity.

The middle term on the right side of (\ref{30.5}) does not appear in
the Chapman-Enskog approach, which is applicable in only the case of very short mean free path.  In that approximation, the system is very close to equilibrium and so (as we shall see in the example below) that extra term would be very small.  But this term is central to the difference of our approximation from the usual first-order results; it is a process-dependent term that allows for
deviations from equilibrium.

To recover the first-order closure approximation found by using the
Chapman-Enskog procedure, we need only neglect our extra term,
which is of order higher than the first since $\mu$ is proportional to $\epsilon$.  But that truncation leads to acausal equations as noted by Israel \cite{isr63, isr76} who (as did others) preferred to go to second order to regain causality.  
 
Another way to avoid the causality problem is to use the moment method based on Grad's approximation \cite{isr79a,isr79b}, but the large number of variables implicated in that procedure makes it hard to apply the results.    Yet the key question is, how good are the results quantitatively in the various approaches?   For answer to this question, we have elsewhere turned to the nonrelativistic case for which empirical tests are available.   These favor the method described here as reported in \cite{dell07,thi03}, for instance.
\subsection{The Fluid Equations}
To complete our statement of the basic equations, we note that
the projections of $P^{\mu\nu}_{\ \ \,,\nu}$ on $u_\mu$ and $h_{\rho\mu}$ are 
\be u_\mu P^{\mu\nu}_{\ \ ,\nu}=\frac{1}{3}\vartheta(E-\epsilon Q)+\Xi^{\mu\nu}_{\ \ ,\nu} u_\mu
\label{32}
\ee  and \be 
h_{\rho\mu}P^{\mu\nu}_{\ \ ,\nu}=\frac{\dot
u^\mu}{3}h_{\rho\mu}(E-\epsilon
Q)-\frac{1}{3}\delta_{\ \rho}^\mu(E-\epsilon
Q)_{,\mu}+h_{\rho\mu}\Xi^{\mu\nu}_{\ \ ,\nu} 
\label{34} \ee 
where 
\be Q=\dot E+\frac{4}{3}E\vartheta \ .\label{33} \ee

On substituting (\ref{32}) and (\ref{34}) into (\ref{28}) and (\ref{29}) respectively, we find \be
Q(1-\frac{1}{3}\vartheta \epsilon)=-u_\mu\Xi^{\mu\nu}_{\ \ ,\nu}
\label{35} \ee 
and, since $h_{\mu \sigma}\dot u^\sigma=\dot u_\mu$,  \be 
\dot u_\mu(4E-\epsilon Q)
=(E-\epsilon Q)_{,\mu}-3h_{\rho\mu}\Xi^{\rho\sigma}_{\ \ ,\sigma} .\label{36} \ee
  
Since (\ref{30.5}) involves a derivative of $E$ as well as $E$ itself, 
it does not have the usual look of a closure relation ---
it is process-dependent.   We may however eliminate $\dot E$ 
at the cost of some complexity.  That is, we could replace $\dot E$ in (\ref{30.5}) by using (\ref{28}).  That would
bring in $u_\mu P^{\mu \nu}_{\ \ ,\nu}\; ,$ which we could replace using (\ref{32}) with (\ref{33}).  That substitution in its turn brings back $\dot E$ but this time as a term ${\mathcal O}(\epsilon^2)$ (since $\mu=P\epsilon$).   If repeated indefinitely, this 
substitution procedure brings in terms of all orders in powers of 
$\epsilon$.  That is the reason that one may expect that (\ref{30.5}) 
can be a significant improvement over the standard closure formulae.   However, we have preferred to leave well enough alone here and to retain the closed form with $\dot E$ rather than the infinite sum.  This sequence of substitutions is different than the 
Chapman-Enskog iteration as there is here no truncation; in any case, we do not use it but merely mention it for clarification.
 
\section{Expansion of a Self-Gravitating Medium}

To provide a simple illustration of the content of the fluid equations
we have just exhibited, we couple them to Einstein's equation and adopt the usual simplifications of elementary cosmology theory --- homogeneity and isotropy.    To avoid wandering too far into complicated (or unknown) physics, we keep to the model of ultrarelativistic, identical particles.   Though this illustration is selected mainly for its lack of distracting details, it is so well 
studied that it makes for good comparisons with earlier work.
We are in fact acting to some extent in the belief
that sufficiently early on in our universe, all particles were ultrarelativistic.   The model we have adopted for the basic fluid 
may not be too misleading in that case.   
\subsection{Two Solutions}
We next study equations (\ref{35}) and (\ref{36}) together with the Einstein field equation \be G^{\mu\nu}= T^{\mu\nu}  \label{37} 
\ee 
where the units are such that the Einstein constant is unity.
Under the conditions of homogeneity and isotropy we have a
conveniently tractable example of flow described by five equations for the five unknowns $E$, $H$ and $u^\mu$.  The configuration then resembles that of the simplest early models of cosmology.   Comparison to the results from that topic will give a first indication of what modification the equations found here may introduce into the study of a self-gravitating medium in the case where the mean flight time of the constituent particles need not be infinitesimal.    

For an isotropic medium with no shear, we have  
$\Xi^{\mu\nu}\equiv 0$ and equations (\ref{35}) and (\ref{36}) become 
\be Q(1-H\epsilon)=0
\label{381}
\ee 
where $H=\vartheta/3$ and 
\be  \dot u_\mu(4E-\epsilon Q)= (E-\epsilon Q)_{,\mu}\ .
\label{39}
\ee 
With $R(t)$ as the scale factor, we have $H=\dot R/R$.

If we adopt the currently preferred cosmological model with 
zero curvature, the Einstein equation (\ref{37}) reduces to
\cite{wei72}
\be  E=3H^2\; 
\label{41}
\ee 
where $E$ is the energy density.  If there were any curvature, the extra term it would introduce would be significant for $R$ close enough to zero.   The rest-mass density has been
left out of account since we consider only the case of a medium dominated by ultrarelativistic particles.    

Equations (\ref{381}), (\ref{39}) and (\ref{41}), admit the following two solutions: 

\begin{enumerate}

\item Solution 1 \\  One solution to (\ref{381}) is $Q=0$.   
In that case (see (\ref{33})), we have
\be \dot E+4E H=0\, ,
\label{42}
\ee 
which gives the familiar result that $ER^4$ is  constant.
If we assume that, as in equilibrium of a photon gas, 
$E\propto T^4$,
we have $T\propto R^{-1}$, the well-known cooling law for equilibrium radiation in an expanding medium.   With $Q=0$, we conclude from (\ref{39}) that $u_\mu \propto R^{-1}$ and we have recovered the features of the standard FRW solution.
 
\item Solution 2 \\  The other possible solution of (\ref{381}) is 
\be H\epsilon = 1. \label{Kn} \ee 
This is really an approximate solution that is good up to errors
of order $\epsilon^2$. The nondimensional parameter, 
$\epsilon H$, is the ratio of the mean free flight time of the particles to the time scale of the macroscopic expansion and is a local Knudsen number in fluid dynamical terminology.
 
Though it is not excluded that we may also have $Q=0$ in this
situation, we shall omit that case here for brevity.  So we
look only at the case where $Q$ is not zero.   The temperature is then not forced to behave in the way that it does in local equilibrium in the FRW model and it is undetermined as yet.

The properties of $\epsilon$ now become significant.   We have already assumed that $\epsilon$  does not depend on $p^\mu$ explicitly, though it may depend on local (in time) macroscopic properties of the medium such as temperature and pressure.  Those variables, in their turns, will generally vary with $R$.   Once all
those dependences are specified, equation (\ref{Kn}) is a differential equation for $R$.  Its solution may be written formally as \be
t = t_0 + \int_{R_0}^R \epsilon \left(T(R)\right)\frac{dR}{R}
\label{soln} \ee
where $R_0=R(t_0)$.  If we specify the relation of $\epsilon$
and $T$, (\ref{soln}) becomes an integro-differential equation for finding $T(R)$.  Then, we would need to solve (\ref{33}) and 
(\ref{39}) for $R(t)$.   

We have not done that as yet as we do not know the nature of
the interactions of the particles in the very early universe .  Rather, 
if as is often done in using the relaxation model, we were to assume for the sake of discussion that $\epsilon$ is constant, $H$ would also be (nearly) constant in this solution.   In that case, we would have \be
R\propto \exp(Ht). \label{infl}\ee
where we have taken $t_0=-\infty$ and $R_0=0$.  Alternatively,
we might have conditions in which solution 1 occurs right after $t=0$ so
that there is a crossover to solution 2 at some $t=t_0$ with appropriate 
matching conditions.  In either case, the e-folding time of $R$ in solution 2 is the mean flight time of the particles.   

Though solution (\ref{infl}) has a certain appeal it does
seem unphysical.  In particular, according to (\ref{41}), constant $H$ 
implies constant $E$.  This awkward outcome originates with the assumption of constant $\epsilon$, which we have introduced only to get a rough idea of the possibilities of solution 2.   
But $\epsilon = 1/\hat\varkappa$ and 
$\hat\varkappa = \hat N\hat\alpha$ where $\hat N$ is the rest frame number density of scatterers and $\hat\alpha$ is the scattering cross-section per scatterer in the rest frame.  In that case, we see that a more reasonable choice is to take
$\hat\alpha$ as constant.  With $\hat N \propto R^{-3}$, we then find that \be
R \propto t^{\frac{1}{3}}, \ee
where we have simplified this solution by choosing $R(0)=0$.    
\end{enumerate}

These possibilities cannot be properly evaluated without a more explicit particle model that we shall not attempt to formulate here.   The main point to be made is that the truncation introduced in the 
Chapman-Enskog procedure has filtered out one of the two solutions for the motion of our extended equations for the 
expansion of a relativistic, self-gravitating gas.   And that second
solution has a different physical origin than the usual ones.

Solution 1 may be considered geometrical in origin; after all, expansion can occur even in a cosmology without matter.  
Solution 2 however is matter-driven as we see from the close connection between the expansion rate and the mean flight time.  
In that sense, solution 2 is analogous to a shock wave whose thickness is typically within an order of magnitude of the mean free path of the particles in the fluid.   But, in soution 2, this shock-like behavior takes place in time rather than in space so that the expansion time scale is analogously of the order of the mean flight time of the constituent particles.  In the two situations, the macroscopic inhomogeneity (whether spatial or temporal) is determined by the microscopic behavior of the particles.   A role of solution 2 in an expansion might be to provide a transition between two different behaviors of solution 1 when there is an interval in which a breakdown of equilibrium occurs.  

\subsection{Entropy Production} 
   
In solution 1, once the medium is in local thermodynamic equilibrium, the expansion will not destroy the equilibrium if the particles are ultrarelativistic.  However, since the expansion rate
and the mean flight time are comparable in solution 2, we cannot
expect any form of thermodynamic equilibrium to be established
in that case.  Yet the relatively infrequent collisions that do occur 
will produce entropy.  The only macroscopic effect in the present
model that can account for this is the volume change which, in
this case, is a disequilibrating effect through (what is known as) the volume viscosity (or bulk viscosity).  In order to gain
some idea whether solution 2 may be of physical interest (if it may
occur at all) we estimate its entropy production rate in order to 
form an impression of what its physical impact on the physical 
state of the gas may be.
 
For the purposes of this brief look at dissipation occurring in 
solution 2 for the expanding medium, we treat $\epsilon$ as constant as we are after only a rough estimate here.   In that case, as we saw already, the
expansion is exponential with $R\propto \exp(Ht)$ and $H$ is
roughly constant.  If this phase of exponential expansion happens early on when $H$ is large, equation (\ref{41}) tells us that the temperature will then be high as well.   

To get some impression of the magnitude of such effects  
we use the formula for the entropy generation rate that has
been computed in the study of radiative fluid dynamics, for
instance in \cite{cherad}.  
There, we find that
\be \dot {\cal S}=\xi\vartheta^2\ee
with \be \xi=\frac{4\epsilon E} {T}\left(\frac{1}{3}+
\frac{\dot T}{T\vartheta}\right)^2\ee
where ${\cal S}$ is the entropy.   
This is a single temperature result that
resembles that found by Weinberg \cite{wei72, ber88} for a two-temperature medium; in each case, $\xi$ vanishes in an FRW equilibrium state.  That is, since 
$\vartheta=3H$, the rate
of entropy generation is 
 \be \dot{\mathcal S}=\frac{4\epsilon E}{T}\left(H+\frac{\dot T}{T}\right)^2,
\label{44} \ee 
which vanishes in solution 1.
 
For this illustration, let us imagine that $\epsilon H$ is small at
the outset and that it grows slowly and passes through unity: the expansion evolves so as to make solution 2 possible.  If a crossover to solution 2 does take place, there will occur a transition period in which  the macroscopic evolution is driven by the interplay of the particle interactions and the large-scale expansion.  
It is internally consistent to take $\epsilon,\ T$ and $H$ as approximately constant before solution 2 gives way to solution 1.    During the time interval when $\epsilon H$ is of the order of unity, solution 2 would then lead to the (approximately constant) entropy generation rate \be
\dot {\cal S} = \frac{4EH}{T}.\ee
In that case, we estimate the entropy at time $t_1$ to be 
 \be {\cal S}= (t_1-t_0) \dot {\cal S}  \; ,
\label{45} \ee 
where solution 2 applies from $t_0$ to $t_1$. 
To find the total comoving entropy we multiply by $R^3$ 
\cite{maa95}.  Then, at the hypothetical instant at
which we presume that solution 2 returns the baton to solution 1, we find that 
  \be 
R^3{\cal S}=R_0^3e^{3H(t_1-t_0)}Q t_1\, .
\label{46} \ee 

To get some idea of what kind of quantitative effect to expect we 
compute the entropy generated during the conventional estimate
of the time interval sometimes adopted for allowing inflation to cure some cosmological ills.  Thus we take the previously suggested 
\cite{kol90} $t_0=10^{-35}$s and ask what the entropy is at
$t_1=10^{-32}$s.  For this, we adopt for $H$ and $T_0$ the values  $6\times 10^{33}$s$^{-1}$ and $10^{28}$K, respectively.  We also have $R_0=ct_0$.  The total comoving entropy generated in the interval $t_0$ to $t_1$ is then of the order 
$10^{88} J/K$.   This estimate is comparable to current 
estimates of the value of the entropy of our universe~\cite{kol90} and has also been 
found when the entropy generation is ascribed to matter/antimatter annihilation \cite{ber88} or to bulk viscosity \cite{maa95} in the standard solution.  Of course, other times and values for the parameters could have been chosen; we have put values found in the literature into our estimates for comparison.   If indeed solution 2 did arise in those very early times it may then have contributed something to the heating of the universe.

\section{Summary}
Already in nonrelativistic kinetic theory, derivations of continuum equations typically give rise to situations in which properties
of the medium are propagated at infinite speed.  Such undesirable
results are produced by the unequal treatments of space and time
in the Chapman-Enskog procedure \cite{uhl63}.    When the same practice is followed in deriving continuum equations in the relativistic case, the C-E procedure again leads to acausal fluid dynamical equations.   Of course, the problem is more urgent in the relativistic case where
the origin of the difficulty has been isolated by Israel \cite{isr63, isr76} and others.    According to  Maartens \cite{maa95},    
\begin{quote}
The problem arises from the first-order nature of the theory, i.e. it considers only first-order deviations from equilibrium. The neglected second-order terms are, in fact, a necessary requirement to prevent non-causal and unstable behavior. These terms transform the equations governing dissipative quantities from the algebraic first-order type into differential evolution equations. A key feature of the second-order theory is that the equilibrium and dissipative variables are considered on the same footing, so that the theory is well suited to dealing with non-stationary processes, such as would occur in the early universe. \end{quote}

The present work is based on the belief that the problem is caused  by the iterative step taken in the C-E procedure, a step that was
likely introduced to make it easier to derive continuum equations  
from Boltzmann's equation.   The point made in the present work is that one should not make the iterative step of Chapman and Enskog in the derivation of the continuum equations.  As the expansion of the one-particle distribution in mean flight time may well not be convergent \cite{gra63}, the iteration that has become standard in such derivations is inappropriate.  On omitting the iterative step of 
Chapman and Enskog, as we have done here, one obtains in first order an approximation for the one-particle distribution function that contains the derivatives that Maartens reports are lost in the C-E procedure.   Formally, those terms are ${\mathcal O}(\epsilon^2)$ in the present work, but it is acceptable to retain them since we are admitting to errors of that order in our first order approximation. 
  
In discussions of these issues one hears the sometimes held opinion that, in a first-order theory, all terms of higher order that appear should be summarily discarded.  But that is not obligatory if the results contain the admission that the error is formally 
${\mathcal O}(\epsilon^2)$.  This error signal does not make it desirable to introduce second order terms {\it ad libidem}, but neither should second-order terms that arise naturally through the
formal development be discarded without a good reason.  And that is the prescription we advocate.

Of course, we have carried out the corrected procedure here for only the relaxation model of relativistic theory, but that is enough to make the case.  (The calculation for the Boltzmann equation is
rather harder and we have so far performed it only for the
nonrelativistic Boltzmann equation with results that will be 
reported elsewhere.)  As was already seen in the nonrelativistic theory, eschewing the C-E iteration does not produce perfection but it does cure the worst problems coming from the C-E procedure as reported in \cite{bgk1,thi03,dell07,chi11}.
  
By starting from the relatively simple relaxation model of kinetic theory, we have been able to bring out the  essential difference between our approach to the derivation of continuum equations
and that of the traditional Chapman-Enskog method.  Though the relaxation model is less specific than the Boltzmann equation as regards the nature of the collisions, it avoids the worrisome features of the relativistic collision process when the particles are not point masses.  Moreover, it does not have the limitation of being confined to binary collisions.   

A possibly interesting sidelight that we have described here is that, when we introduce the Einstein equation together with the usual simplifications of elementary cosmological models, we find a second   expanding solution in addition to the usual one of
the FRW model for the case of ultrarelativistic particles such as
photons.   In this second solution, the time scale of the
global expansion is of the order of the mean flight time of the
constituent particles.   This feature of the temporal behavior is
somewhat reminiscent of classical shock waves whose spatial extent is comparable to the mean free paths of the particles of the medium.   Analogously, solution 2 has an expansion rate comparable to the mean flight time in the particle interactions.  

In turning to cosmology for our illustrative example of an
expanding medium we felt that, as Weinberg has put it \cite{wei72}, ``the temptation to try to construct some sort of model of the very early universe is irresistible."   In any case, whether 
there is any relevance to cosmology or not in our look at an 
expanding medium, the simplifications of elementary cosmology theory have at least provided us with a tractable illustrative example.   

\bibliographystyle{prsty}

\end{document}